\documentclass[%
 reprint,
superscriptaddress,
 amsmath,amssymb,
prl,
]{revtex4-1}

\usepackage{graphicx}
\usepackage{dcolumn}
\usepackage{bm}

\usepackage{braket}
\usepackage{color}

\begin{document}

\preprint{APS/123-QED}

\title{{Strain-Tunable GaAs Quantum dot: A Nearly Dephasing-Free Source of Entangled Photon Pairs on Demand}}

\author{Daniel Huber}
\email{daniel.huber@jku.at}
  
\author{Marcus Reindl}%
    \thanks{These two authors contributed equally}
	\affiliation{Institute of Semiconductor and Solid State Physics, Johannes Kepler University, Linz, Altenbergerstr. 69, 4040, Austria}

\author{Saimon Filipe Covre da Silva}%
    \thanks{These two authors contributed equally}
	\affiliation{Institute of Semiconductor and Solid State Physics, Johannes Kepler University, Linz, Altenbergerstr. 69, 4040, Austria}
       
\author{Christian Schimpf}%
\affiliation{Institute of Semiconductor and Solid State Physics, Johannes Kepler University, Linz, Altenbergerstr. 69, 4040, Austria}

\author{Javier Mart\'{i}n-S\'{a}nchez}%
	\affiliation{Institute of Semiconductor and Solid State Physics, Johannes Kepler University, Linz, Altenbergerstr. 69, 4040, Austria}
    \affiliation{Department of Physics, University of Oviedo, 33007 Oviedo, Spain}
    
 \author{Huiying Huang}%
	\affiliation{Institute of Semiconductor and Solid State Physics, Johannes Kepler University, Linz, Altenbergerstr. 69, 4040, Austria}
    
\author{Giovanni Piredda}%
	\affiliation{Forschungszentrum Mikrotechnik, FH Vorarlberg, Hochschulstr. 1, A-6850 Dornbirn, Austria}
    
\author{ Johannes Edlinger}%
	\affiliation{Forschungszentrum Mikrotechnik, FH Vorarlberg, Hochschulstr. 1, A-6850 Dornbirn, Austria}
    
\author{Armando Rastelli}%
\email{armando.rastelli@jku.at}
	\affiliation{Institute of Semiconductor and Solid State Physics, Johannes Kepler University, Linz, Altenbergerstr. 69, 4040, Austria}

\author{Rinaldo Trotta}%
\email{rinaldo.trotta@uniroma1.it}
	\affiliation{Institute of Semiconductor and Solid State Physics, Johannes Kepler University, Linz, Altenbergerstr. 69, 4040, Austria}
    \affiliation{Department of Physics, Sapienza University of Rome, Piazzale Aldo Moro 5, 00185 Rome, Italy}





\begin{abstract}
We report on the observation of nearly maximally-entangled photon pairs from semiconductor quantum dots, without resorting to post-selection techniques. We use GaAs quantum dots integrated on a patterned piezoelectric actuator capable of suppressing the exciton fine structure splitting. By using a resonant two-photon excitation we coherently drive the biexciton state and demonstrate experimentally that our device generates polarization-entangled photons with a fidelity of 0.978(5) and a concurrence of 0.97(1) taking into account the nonidealities stemming from the experimental setup. By combining fine-structure-dependent fidelity measurements and a theoretical model, we identify an exciton spin-scattering process as a possible residual decoherence mechanism. We suggest that this imperfection may be overcome using a modest Purcell enhancement so as to achieve fidelities $>$0.99, thus making quantum dots evenly matched with the best probabilistic entangled photon sources. \end{abstract}

\pacs{Valid PACS appear here}
\maketitle


For the implementation of quantum computation and communication protocols, highly entangled photons are a fundamental building block \cite{Kimble:Nat2008,Zeilinger2017}. So far, the state of the art sources for the generation of near maximal entangled photon states are based on parametric-down-conversion (PDC) processes, where fidelities larger than 0.99 have been reported ~\cite{Weston2002,PhysRevLett.115.250401,PhysRevLett.115.250402,Jons2017}. However, such sources of entangled photons are not ideal for quantum communication protocols due to lack of on-demand emission. A potential solution to this hurdle is provided by semiconductor quantum dots (QDs), which can generate pairs of polarization entangled photons via the biexciton (XX)-exciton (X) cascade~\cite{PhysRevLett.84.2513,PhysRevLett.96.130501,1367-2630-8-2-029}. This approach is promising for applications, not only because QDs are compatible with current photonic integration technologies, but in particular because entangled photons can be generated on-demand~\cite{Michler:NaturePhoton2014}, with high efficiency, and with high degree of indistinguishability~\cite{Senellart:NatPhoton2016,PhysRevLett.116.020401}. Up to now, the on-demand photon-pair preparation with near-unity degree of entanglement has remained elusive. This hurdle is related to the presence of several decoherence mechanisms typical of the solid state system. The most prominent obstacle is related to the presence of an energy splitting between the two intermediate X states, the so-called fine structure splitting (FSS)~\cite{Bayer:PRB2002}. Strictly speaking, a static FSS is not a source of decoherence \textit{per se}, but leads to an evolution of the entangled state over time according to~\cite{Stevenson:PRL2008}:
\begin{equation}
      \ket{\psi}=1/\sqrt{2}(\ket{H_{\text{XX}}}\ket{H_{\text{X}}}+e^{\frac{i S t}{\hbar}}\ket{V_{\text{XX}}}\ket{V_{\text{X}}}),
			\label{eq:bellState_fss}
\end{equation}
where $S$ is the FSS, $t$ the time between XX and X photon emission and $H_{XX}$ ($V_{XX}$) and $H_{X}$ ($V_{X}$) are XX and X photons in the linear horizontal (vertical) polarization base, respectively. It is obvious that in presence of a FSS the time-averaged fidelity to an entangled Bell states is determined by the temporal resolution of the experimental setup as compared to the exciton lifetime $\tau_{1}$. A possible way to circumvent this problem is temporal post-selection~\cite{doi:10.1021/nl503581d,Ward2014,Salter2010,PhysRevApplied.8.024007} that, however, lowers the effective brightness of the source. Alternatively, external optics could be used to “compensate” for the evolving character of the entangled state~\cite{PhysRevB.95.235435,APL201103,Fognini2017}. Nonetheless, the need for complex and bulky optics in combination with post-selection techniques makes QDs less appealing for scalable quantum technologies. It is evident from Eq.~\ref{eq:bellState_fss} that all these complications could be avoided using QDs with suppressed FSS. Among the different ways to reduce/suppress the FSS (see Ref.~\cite{Plumhof2012,PhysRevLett.109.147401,martin2017strain}), the one that exploits triaxial strain-tuning~\cite{Trotta:2016NatCom} is probably the most promising, as it can be used to fine-tune the FSS of arbitrary QDs to zero and also to set the emission energy to predefined values. Yet, experiments have shown that even at zero FSS the degree of entanglement is still far from being optimal~\cite{Trotta:NanoLet2014,Trotta:2016NatCom,Chen2016}. This has highlighted the existence of additional dephasing mechanisms, most notably (i) recapture~\cite{Dousse:Nat2010,PhysRevB.88.041306,PhysRevLett.102.030406} and (ii) X spin-flip processes~\cite{Chekhovich2011,PhysRevLett.99.266802}. (i) is related to re-excitation of the  intermediate X level to the XX level before its decay to the ground state. This effect can be avoided using two-photon resonant excitation~\cite{PhysRevLett.110.135505,Michler:NaturePhoton2014} that, in turn, ensures on-demand generation of entangled photons. (ii) is instead believed to arise from the interaction between the nuclear spin ensemble and the X as well as from scattering with excess charges~\cite{MichlerQDs,Fognini2017}. While the role of the nuclear spins is questionable~\cite{PhysRevB.70.033301,Fognini2017}, experiments performed with In-free QDs driven resonantly  have indeed shown unprecedented, albeit not-yet optimal, levels of entanglement~\cite{Huber2017,Keil2017,Reindl:nano2017,Kuroda:2013PRB,doi:10.1021/acs.nanolett.7b04472}. Since all these experiments have been performed in QDs with non-zero FSS, it remains unclear whether a QD can be really considered as a dephasing-free entanglement photon source and, most importantly, whether near maximal entangled photons can be  experimentally achieved without resorting to post-selection. In order to answer these questions, we perform quantum state tomography of photons emitted by strain-tunable GaAs QDs with suppressed FSS and driven under two-photon resonant excitation. We demonstrate that our source can generate photon-pairs with a high fidelity (concurrence) of 0.978(5) (0.97(1)) without the need of post-selection techniques. Although these values are the highest ever measured with a QD emitter~\cite{Huber2017,Keil2017,PhysRevApplied.8.024007}, we still observe a small, but significant deviation from the case of a maximally entangled state. In order to investigate this deviation in more detail, we measured the degree of entanglement against the FSS and use the model proposed in Ref.~\cite{PhysRevLett.99.266802} to determine the origin of the residual decoherence mechanisms. Our calculations show that the deviation can be explained by the presence of a remaining exciton spin scattering process, whose impact can be alleviated using photonic structures enabling a modest Purcell enhancement. 

\begin{figure}[htbp]
	\centering
		\includegraphics[width=90mm]{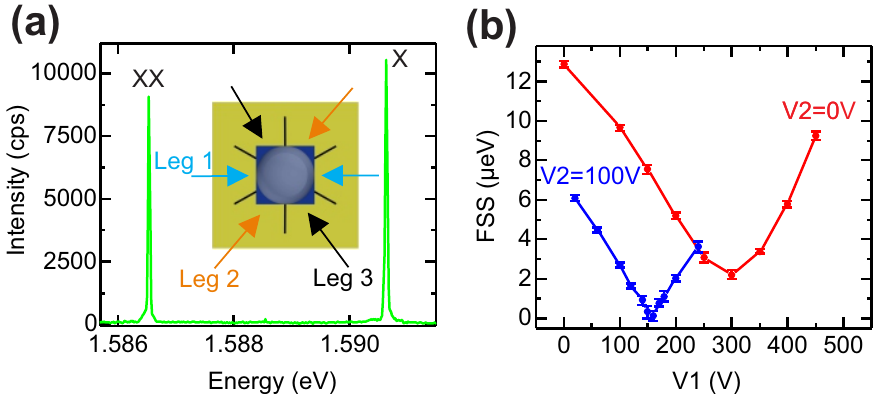}
	\caption{\textbf{Erasure of the fine-structure splitting via a strain-tunable device.} (a) Spectrum of a representative two-photon resonant excited GaAs quantum dot. The inset shows a sketch of the used 6-leg device from the top. The sample (blue) with a solid immersion lens on top is bonded onto the piezo-electric actuator (golden part). The piezo is structured using three cuts (black areas) into six legs which are pairwise electrically connected on the backside (Leg 1-3). (b) Minimization of the  FSS as described in the text by tuning the voltage on Leg 1 for V2=0 (red) and V2=100V (blue), respectively.} 
	\label{fig:fig2}
\end{figure}

The GaAs QDs - fabricated via Al droplet etching via molecular beam epitaxy at JKU Linz - are embedded in a planar distributed Bragg reflector cavity for increasing the photoluminescence intensity. The sample substrate is thinned down to a 30 $\mu$m thick micro-membrane, which is bonded on top of a micro-machined [Pb(Mg$_{1/3}$Nb$_{2/3}$)O$_{3}$]$_{0.72}$-[PbTiO$_{3}$]$_{0.28}$ piezoelectric actuator \cite{sanchez:AM2016,Trotta:2016NatCom,Ziss2017} (for details on the sample and device fabrication see supplementary Sec. I A). The device features six areas separated by air gaps, the so called legs (see inset Fig. \ref{fig:fig2} (a)). The legs are aligned at an angle of 60$^\circ$ with respect to each other and opposite legs are pairwise electrically connected. The three resulting leg pairs (labeled as Leg 1,2,3) are isolated from each other. The design allows three independent quasi-uniaxial stresses to be applied in the membrane plane by setting three independent voltages (labeled as V1,V2,V3) at the bottom of the Legs 1-3 with respect to the gold coated topside of the piezoelectric actuator, which acts as a ground contact. It is well known that two external fields with independent degrees of freedom are required to cancel the FSS in a QD with an arbitrary anisotropy in the confinement potential \cite{PhysRevLett.114.150502,PhysRevLett.109.147401}. In our case, we use two legs of the piezoelectric actuator for this purpose. For the full capabilities of the device structure we refer the interested reader to \cite{sanchez:AM2016,Trotta:2016NatCom}. For the experiment discussed below we select an arbitrary QD and resonantly pump the XX cascade via a two-photon excitation with a $\pi$-pulse. The resulting emission spectrum is shown in Fig. \ref{fig:fig2} (a). We determine an initial FSS of 12.9(2) $\mu$eV, which is a large value for this type of QD \cite{Huo:APL2013}. Yet, the six-legs device can tune the FSS to zero using Leg 1 and Leg 2 only. In fact, by tuning V1 (see red curve in Fig. \ref{fig:fig2} (a)) one can see that the FSS decreases, reach a minimum ($S \neq 0$) and increases again. This is an expected behavior, as the polarization direction of the X emission at zero applied voltage differs from the direction of the stress exerted by Leg 1. In order to suppress the FSS it is sufficient to first use a second leg (here: Leg 2) to align the QD anisotropy in the stress direction of Leg 1 (by setting V2=100 V) and then tuning Leg 1 to find the minimum FSS. As shown by the blue curve in Fig. \ref{fig:fig2} (b) the procedure allows us to tune the FSS to 0.1(2) $\mu$eV, a value which is below the spectral resolution of our measurement system. 

We now measure the degree of polarization entanglement of the photons emitted by the XX-X cascade at zero FSS. Therefore, we reconstruct the two-photon density matrix (DM) by performing polarization resolved cross-correlation measurements between X and XX photons. To spectrally separate the X and XX lines and to remove scattered laser light as well as background emission, we use a set of volume Bragg gratings as described in more detail in supplementary Sec. I B. However, such filters as well as other components of the setup can introduce a rotation in the polarization state of the emitted photons. Such a rotation, which does not lower the degree of entanglement itself, can reduce the fidelity to the expected Bell state $\ket{\psi^+}$, a parameter which is of crucial importance when it comes to potential applications like quantum teleportation. Therefore, we take special care of the polarization response of our setup and use a set of variable liquid crystal retarders to compensate for any unitary polarization-rotation introduced by the experimental apparatus. Moreover, we  fine-tune the polarization compensation by minimizing the coincidences between right (left) circular polarized XX and right (left) circular polarized X photons\cite{Versteegh:NatCom2014}. 
\begin{figure}[htbp]
	\centering
		\includegraphics[width=90mm]{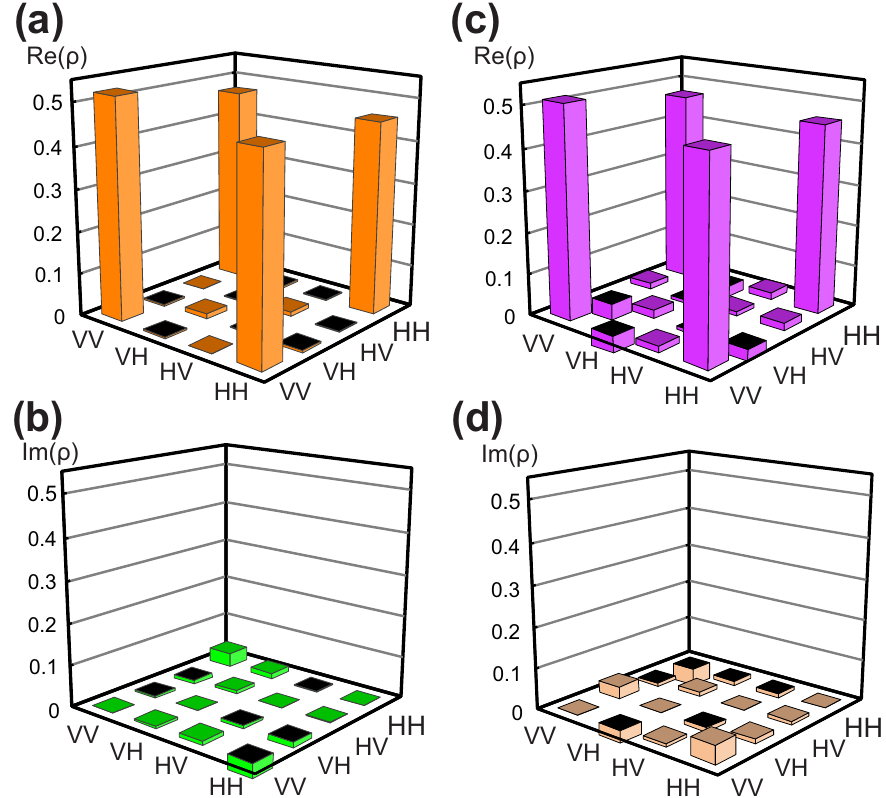}
	\caption{\textbf{Two-photon density Matrix of two representative GaAs quantum dots at zero fine-structure splitting.} Real (a) and imaginary part (b) of the measured two-photon density matrix for QD1 at zero fine structure splitting. (c) and (d) same measurement as in (a) and (b), but with a different quantum dot (QD2).} 
	\label{fig:fig3}
\end{figure}
The resulting DM for the selected QD (QD1) as obtained by a set of 36 correlation measurements with the aid of a maximal likelihood method \cite{White:PRA2001} can be seen in Fig. \ref{fig:fig3} (a) and (b). The resulting fidelity with respect to the $\ket{\psi^+}$ state is $f=$0.960(2) with a highest eigenvalue of $e=$0.962(3). Further, we calculated the concurrence to $\zeta=$0.922(5), which also indicates a high degree of entanglement. It is also worth mentioning that using a reduced measurement set of only 6 correlation measurements (see supplementary Sec. II) to calculate the fidelity according to:
\begin{equation}   
        f=\frac{1+C_{\text{linear}}+C_{\text{diagonal}}-C_{\text{circular}}}{4},
	 \label{eq:fid_simple}
\end{equation}   
(where $C$ are the correlations visibilities) gives a similar fidelity of $f=$0.959(7). All the errors within this work given for the fidelity, concurrence and eigenvalue are calculated by using Gaussian error propagation and/or Monte Carlo method assuming a Poisson distribution of the measured coincidence counts. In order to confirm the generality of our results, we repeated the study on a second, randomly selected QD (QD2) (see  Fig. \ref{fig:fig3} (c) and (d)) and obtained $f=0.953(2)$, $e=0.960(2)$ and $\zeta=0.919(4)$. Such an unprecedented level of entanglement already allows for quantum communication applications, as error correction protocols can compensate for the residual imperfections \cite{PhysRevA.66.060302}. However, here we are interested in answering the following questions: What is preventing the degree of entanglement to be ideal? And, most importantly, can QDs be really considered as decoherence-free entangled photon sources? While previous works have theoretically suggested that the answer to the latter question is positive\cite{Fognini2017}, an experimental demonstration of near-maximally entangled photons from QDs is still lacking. 

\begin{figure}[htbp]
	\centering
		\includegraphics[width=90mm]{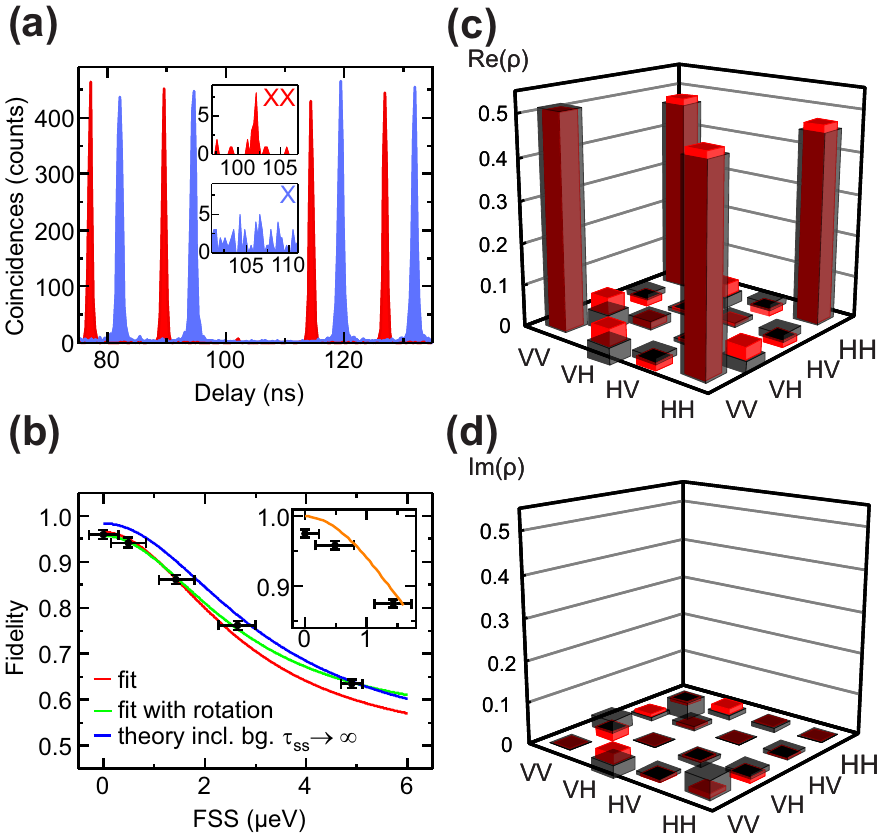}
	\caption{\textbf{Near-maximally entangled photons from quantum dots.} (a) Autocorrelation measurement of the biexciton (XX) (see red curve) and exciton (X) (see blue curve) from QD1. To improve readability the XX and X the curves are shifted by 5 ns. (b) Entanglement fidelity versus fine structure splitting (S) for QD1. The black data points represent the measurement data. The red curve is a fit according to Eq. \ref{eq:fid_fit}, while the green curve is a fit taking into account a rotation of the state, as explained in the text. The blue curve takes into account background laser-light (bg) but no spin dephasing $(\tau_{ss}\rightarrow\infty)$. The inset shows the evolution of the fidelity versus the FSS for background-free entangled photons and no spin dephasing (orange line). The data points (black) are the measured data from the main figure, but corrected for the estimated background. Real (c) and imaginary part (d) of the two-photon density matrix measured on QD2. The gray bars show the results calculated out of the raw data, while the red bars are related to the matrix after the correction for background photons, waveplate retardance and dark counts.} 
	\label{fig:fig4}
\end{figure}

In order to answer these questions, we first have a closer look at the experimental setup. We identify three sources of errors: (i) The detector dark counts, (ii) the retardance of the wave plates used for the reconstruction of the DM, and (iii) background photons. Subtracting the dark counts leads to a 0.3$\%$ improvement for the fidelity and 0.8$\%$ for the concurrence for both measured QDs (the dark count rate of our detector is $<20$ Hz). (ii) The retardance of the waveplates is instead a more delicate issue. According to the formalism presented in Ref. \cite{White:PRA2001}, a tomographically complete measurement set is required for the calculation of the DM. Due to imperfections of the waveplates, used to project the two-photon state into the different bases, the real measurement base will deviate from the one assumed in the calculation. Therefore, we incorporate in the calculation the real retardance of our achromatic waveplates at the emission wavelength of the QD into the computation with 0.516 waves and 0.258 waves (according to the data sheet provided by the constructor) for the lambda/2 and lambda/4, respectively. The position accuracy of the fast axis is $0.02^\circ$ and thus negligible (for details see supplementary Sec. III). Taking into account the dark counts and the effect of the wave plates, the imaginary elements $\bra{HH}\rho\ket{VV}$ and $\bra{VV}\rho\ket{HH}$ of the DM shown in Fig. \ref{fig:fig3} disappear and the resulting values for fidelity and concurrence are $f=0.968(2)$ and $\zeta=0.936(5)$ and $f=0.958(2)$ and $\zeta=0.925(5)$ for QD1 and QD2, respectively. Finally, we investigate the effect of (iii) by measuring the $g^{(2)}$ autocorrelation function for XX and X photons. For QD1 (QD2) we measure a value of $g_{XX}^{(2)}(0)=0.014(3)$ ($g_{XX}^{(2)}(0)=0.021(5)$) and $g_{X}^{(2)}(0)=0.008(2)$ ($g_{X}^{(2)}(0)=0.015(3)$). These values are related to the excitation laser, as similar experiments recently performed on the same QDs (but using polarization suppression to reject laser light) provide values of $g^{(2)}(0)$ which are orders of magnitude smaller \cite{Schweickert2017}. In order to support this statement, we performed additional autocorrelation measurements for XX and X in all the polarization bases (see supplementary Sec. IV) needed to reconstruct the DM and found that the background is primarily linearly (vertically) polarized. On the one hand, this confirms that the background photons originate from the excitation laser. On the other hand, with the help of a statistical model (see supplementary Sec. VI) these $g^{(2)}$ measurements can be used to correct the DM. This is shown for QD2 in Fig. \ref{fig:fig4} (c) and (d), where one can see the effect of background photons. We calculated the fidelity out of the corrected DM and found $f=0.978(5)$, which is an increase of $2.6\%$. The largest eigenvalue improved to $e=0.981(5)$ and the concurrence is $\zeta=0.97(1)$. From a fundamental point of view it is interesting to check whether there is a remaining decoherence mechanism occurring during the cascade decay. To do so, we investigate the degree of entanglement as a function of the FSS, as detailed below.

We start out by considering a possible residual FSS $S_0$=250~neV (corresponding to the resolution of the used setup), and a background according to the $g^{(2)}$ measurements discussed above. Additional FSS fluctuations are expected because of the fluctuating Overhauser field $B_{OH}(t)$~\cite{RevModPhys.85.79}. Since our FSS measurements are performed on timescales of seconds, we cannot quantify such fluctuations experimentally. To estimate their amplitude we assume a maximum field of $B_{max}=4$ T~\cite{Vandersypen:NatMat2013} with a standard deviation of $\sigma=B_{max}/\sqrt{N}\approx 6 $ mT, where $N$ is the number of spin-3/2 nuclei in the QD material ($N\approx 4\cdot 10^5$ for our QDs). For QDs, the  effect of the Overhauser field on the FSS is dominated by its vertical component ($z$)~\cite{Stevenson:arXiv2011}, so that $S=S_0+\mu_{B}(g_{e,z}+g_{h,z}) B_{OH,z}(t)$~\cite{Bayer:PRB2002}, with electron (heavy hole) g-factor $g_{e,z}=-0.15$ ($g_{h,z}=1.1$) according to Ref.~\cite{PhysRevB.93.165306}. With these assumptions we find out that the measurement data for QD1 (QD2) still deviate by 4.5 (2.4) standard deviations from theory (provided by the state in Eq. \ref{eq:bellState_fss}). This result shows that an additional dephasing mechanism is at play. To verify its impact we make use of the spin-scattering model presented in Ref. \onlinecite{PhysRevLett.99.266802} and investigate the fidelity versus the FSS for QD1:
\begin{equation}   
      f=\frac{1}{4}(1+k g + \frac{2 k g}{1+(\frac{g S \tau_{1}}{\hbar})^2}).
	 \label{eq:fid_fit}
\end{equation}
Here $k$ is the proportion of the light exclusively emitted by the QD and $g=\frac{1}{1+\tau_{1}/\tau_{SS}}$ the fraction of photons not influenced by spin scattering with $\tau_{ss}$ the characteristic X spin scattering time. To reduce the number of free parameters in Eq. \ref{eq:fid_fit}, we measured the lifetime in a fluorescence decay experiment and found $\tau_{1}=241(10)$ ps. Further, using statistical considerations we estimate $k$ to be: 
\begin{equation}   
      k\approx 1-g_{X}^{(2)}(0)-g_{XX}^{(2)}(0)+g_{X}^{(2)}(0)g_{XX}^{(2)}(0)=0.978(4).
	 \label{eq:kval}
\end{equation}
 The fit of the measurement data is presented in Fig. \ref{fig:fig4} (b) (red curve), which yields a value for the spin dephasing of $\tau_{ss}=11(8)$ ns. In addition to the fit also the theoretical curve without the presence of spin dephasing ($\tau_{ss}\rightarrow \infty$ ) but in presence of the measured laser-photons background is plotted (see blue curve). The latter one (theoretical curve for decoherence-free entanglement) shows a larger deviation at small FSS, while the former (fit) reveals a deviation at FSS $>2$ $\mu$eV. The deviation between fit and measurement can be explained by the fact that the fidelity is only estimated using the correlation visibilities $C$ in the linear, diagonal, and circular base (see Eq. \ref{eq:fid_simple}). In case of zero FSS the DM as well as Eq. \ref{eq:fid_simple} yield the same fidelity. However, this does not hold if the FSS $\neq 0$ and the entangled state contains an additional phase factor $\omega$ introduced by the measurement setup (see supplementary Sec. V). If we include $\omega$ in the fitting routine (see green curve) we obtain $\omega=-9(4)^{\circ}$ and $\tau_{ss}=14(10)$ ns. The large error of $\pm 10$ ns does not allow us to draw a definite conclusion about the origin of the spin scattering. A plausible explanation is the interaction between the confined exciton and charges in the vicinity of the QD~\cite{MichlerQDs,Fognini2017}. If we use Eq. \ref{eq:fid_fit} to estimate a background correction for the datapoints in Fig. \ref{fig:fig4} the fidelity at FSS$=0$ shows a significant deviation from the ideal case (see inset Fig. \ref{fig:fig4}). By considering the measured X lifetime of QD 2, which is $\tau_{1}=290(5)$ ps, and the fitted spin dephasing time, we can estimate the highest achievable fidelity using Eq. \ref{eq:fid_fit}, with $S=0$ and $k=1$ to f=0.98(1), which is within the error of the corrected fidelity presented above.  

In summary, our results show that, by canceling the FSS, our strain-tunable QDs can generate nearly-maximally entangled photons pairs on demand. By looking at the concurrence (fidelity), the level of entanglement reported here represents a 10\% (4\%) increase as compared to the best QD source of entangled photons reported to date~\cite{Huber2017}.
Further, even with temporal post selection such a high degree of fidelity has not been observed so far \cite{PhysRevApplied.8.024007}. However, the data indicate the presence of an almost-negligible, albeit non-zero, decoherence mechanisms, likely related to spin-scattering. Nevertheless, we suggest the use of a photonic structure would allow this problem to be overcome. In particular, by increasing the Purcell factor from $\approx 1$ in the used device to 3 - a value which may be achieved in photonic structures compatible with non-degenerate entangled photon generation~\cite{Dousse:Nat2010} - the expected entanglement fidelity would surpass $0.99$ and lift QD entanglement to the same level as PDC \cite{Weston2002,PhysRevLett.115.250401,PhysRevLett.115.250402,Jons2017}. It is also worth to mention that, differently from previous works \cite{Trotta:2016NatCom,sanchez:AM2016}, the device reported here uses membranes with a thickness of 30 microns instead of few hundred nanometers. Such an approach is compatible with the processing steps required to fabricate state of the art photonic structures \cite{Nowak2014,sapienza2015nanoscale} and would allow for boosting the flux of photons so as to realize the ideal source of entangled photons needed for quantum communication. 

\section*{Acknowledgement}

This work was supported by the Austrian Science Fund (FWF):
P29603, the European Research Council (ERC) under the European Unions Horizon 2020 research and innovation programme (SPQRel, Grant Agreement 679183), the AWS Austria Wirtschaftsservice, PRIZE  Programme, under
Grant No. P1308457. We thank G. Weihs, B. Pressl, and R. Keil (Univ. Innsbruck) and V. Volobuev, Y. Huo for fruitful discussions and U. Kainz for help during the device fabrication.

\newpage

\bibliography{references}

\end{document}